\documentclass[a4paper,12pt]{article}
\usepackage{authblk}
\usepackage{amsmath,amssymb,amsfonts,amsthm}
\usepackage{graphicx}
\usepackage{subfigure}
\usepackage{geometry}

\newcommand{\ie}{{\it i.e.}\ }
\newcommand{\eg}{{\it e.g.}\ }
\newcommand{\etal}{{\it et al.}\ }

\title{A one-phase Stefan problem with size-dependent thermal conductivity}

\author{F. Font} 
\affil{Department of Physics, Universitat Polit\`{e}cnica de Catalunya, Av. Dr. Mara\~{n}on 44-50, 08028 Barcelona, Spain}

\begin{document}

\maketitle

\begin{abstract}
In this paper a one-phase Stefan problem with size-dependent thermal conductivity is analysed. Approximate solutions to the problem are found via perturbation and numerical methods, and compared to the Neumann solution for the equivalent Stefan problem with constant conductivity. We find that the size-dependant thermal conductivity, relevant in the context of solidification at the nanoscale, slows down the solidification process. A small time asymptotic analysis reveals that the position of the solidification front in this regime behaves linearly with time, in contrast to the Neumann solution characterized by a square root of time proportionality. This has an important physical consequence, namely the speed of the front predicted by size-dependant conductivity model is finite while the Neumann solution predicts an infinite and, thus, unrealistic speed as $t\rightarrow0$. 
\end{abstract}

\section{Introduction}


The Stefan problem, describing the phase change of a material, is one of the most popular problems in the moving boundary problem literature. Typically, it requires solving heat equations for the temperature in the two phases (\eg solid and liquid), while the position of the front separating them, the moving boundary, is determined from an energy balance referred to as the Stefan condition. The Stefan problem has been studied in great detail since Lam\'{e} and Clapeyron formulated it in the 19th century \cite{Lam31}. There are several reference books that the reader may consult for a comprehensive background on the classical problem \cite{Rub,Crank,Hill87,Alex93,Cars,Gupta}. The problem only admits an exact solution in the Cartesian one-dimensional case, referred to as the Neumann solution (see, for instance, \cite{Cars}). Solutions in other geometries  or higher dimensions are usually found numerically or via asymptotic/perturbation techniques (see, for example, \cite{McCue2055} for an investigation of the classical two-phase Stefan problem in a sphere or an analysis of the solidification of a liquid half-space \cite{Wallman1397}). A complete literature review of the classical Stefan problem is not the purpose of this study and the reader is referred to the new edition of the book by Gupta \cite{Gupta} for an up to date bibliography on the problem. 

The classical problem has been modified in different ways to introduce new physical phenomena such as supercooling or curvature dependent phase change temperature \cite{Xie90,evans00,WU20092349,Font2013,Font2014,Bac14,Drag16,Cho74}. Modifications are usually linked with a thermophysical property of the material which changes with, for example, the geometry of the system (\eg curvature-induced melting point depression \cite{WU20092349,Font2013,Font2014,Bac14,Drag16}), the speed of the moving boundary (\eg supercooling \cite{Xie90,evans00,Font2013a}), or the temperature itself \cite{Cho74}. 
Motivated by recent experimental studies on Silicon nanofilms and nanowires showing that the thermal conductivity decreases as the size of the physical system decreases \cite{Li03,Liu04}, in this work we will consider the effect of size-dependent thermal conductivity on the solidification process of a one-dimensional slab. In Jou \etal \cite{JOU2005963} an analytical expression for the dependence of the thermal conductivity of a solid on the size of the physical system is derived from the Extended Irreversible Thermodynamics theory \cite{Joubook}. Assuming that all the phonon mean-free paths and relaxation times are equal, their expression for the thermal conductivity takes the form
\begin{align}\label{eq1}
k = \frac{2k_0 L^2}{l^2} \left(\sqrt{1+\frac{l^2}{L^2}}-1\right)\,, 
\end{align}
where $k_0$ represents the bulk thermal conductivity of the solid, $L$ the size of the solid, $l$ the phonon mean free path  \cite{JOU2005963}. In a follow-up paper, expression \eqref{eq1} is tested against experimental data showing good agreement \cite{Alv07}. The main goal of this paper is to study the effect of the size-dependent thermal conductivity on a solidification process by introducing \eqref{eq1} in the formulation of the one-phase Stefan problem.  

The paper is organized as follows. In the next section we formulate the Stefan problem with size-dependent thermal conductivity and discuss how the Neumann solution can be retrieved. In section 3 we provide a perturbation solution based on large Stefan number. In section 4 we present the numerical strategy to solve the problem and analyse the small time limit, which is needed to initialize the numerical scheme. In section 5 we discuss our results and in section 6 we draw our conclusions. 

\section{Problem formulation}

Consider a liquid initially at its equilibrium freezing temperature, $T_f$, occupying the space $x\geq0$. Suddenly, the temperature is lowered to $T_c<T_f$ on the edge $x=0$ and the liquid starts to solidify. The newly created solid phase will start to grow occupying the space $0<x<s(t)$, where $s(t)$ represents the position of the solidification front as well as the size of the solid phase. The temperature of the solid phase, $T(x,t)$, is described by
\begin{align}\label{dim1}
\rho c \frac{\partial T}{\partial t} &= k(s(t)) \frac{\partial^2 T}{\partial z^2} \qquad \mbox{on} \qquad 0<z<s(t) \,,
\end{align}
where $\rho$ is the density, $c$ the specific heat, and $k(s(t))$ the thermal conductivity which depends on the size of the solid phase and is obtained by setting $L=s(t)$ in \eqref{eq1}. That is, 
\begin{align}\label{eq2}
k(s(t)) = \frac{2k_0 s(t)^2}{l^2} \left(\sqrt{1+\frac{l^2}{s(t)^2}}-1\right)\,. 
\end{align}
Note, in the current study, we consider the one-phase approximation, \ie we assume the liquid to be at the equilibrium freezing temperature, so no equation is needed for the temperature of the liquid phase. The temperature of the solid is subject to the boundary conditions 
\begin{align}
T(0,t) = T_c\,,\qquad T(s(t),t) = T_f\,. 
\end{align}
Finally, at the solidification front we have the Stefan condition 
\begin{align}\label{dim4}
\rho\,\Delta H\,\frac{ds}{dt} = k(s(t)) \left.\frac{\partial T}{\partial z}\right|_{z=s(t)} \,, 
\end{align}
where $\Delta H$ is the latent heat of fusion. To obtain physically meaningful results to our problem we use the parameter values for Silicon in Table \ref{table1}.     

\begin{table}
\begin{center}
\begin{tabular}{ |c|c|c|c|c|c|}
\hline
\hline
$\rho$ (kg/m$^3$) & $c$ (J\,/kg$\cdot$K) & $k_{0}$ (W/m$\cdot$K) & $T_{f}$ K & $\Delta H$ (J/kg) & $l$ (m) \\
\hline
2320    & 770 & 120     & 1687    & 1926$\times10^{3}$  & 40$\times10^{-9}$\\ 
\hline
\hline
\end{tabular}
\caption{Thermophysical properties of Silicon.}
\label{table1}
\end{center}
\end{table}

\subsection{Nondimensional model} 

Introducing the nondimensional variables  
\begin{align*}
T^* = \frac{T - T_f}{T_f-T_c}\,, \qquad x^* = \frac{x}{l} \,, \qquad t^* = \frac{k}{l^2 \rho c}\,t \,, \qquad s^*= \frac{s}{l} \,, 
\end{align*}
in \eqref{dim1}-\eqref{dim4} and removing the star notation, we obtain the dimensionless model 
\begin{subequations}
\begin{align}
\frac{\partial T}{\partial t} &= 2s\left(\sqrt{s^2+1}-s\right) \frac{\partial^2 T}{\partial z^2} \qquad \mbox{on} \qquad 0<z<s(t) \,, \label{eq_non_1}\\
T(0,t) &= -1\,,\\
T(s(t),t) &= 0\,,\\ 
\beta \frac{ds}{dt} &= 2s\left(\sqrt{s^2+1}-s\right) \left.\frac{\partial T}{\partial z}\right|_{z=s} \,, \label{eq_non_4}\\ 
\quad s(0)&=0\,, \label{eq_non_5}
\end{align}
\end{subequations}
where $\beta = \Delta H/\rho c(T_f-T_c)$ is the Stefan number (ratio of latent heat to sensible heat). The term $2s\left(\sqrt{s^2+1}-s\right)$ in \eqref{eq_non_1}-\eqref{eq_non_5} represents the thermal conductivity \eqref{eq2} in dimensionless form, which takes the values 0 for $s=0$ and $1$ for $s\rightarrow\infty$.

\subsection{Neumann solution}

The classical one-phase Stefan problem is retrieved by setting $2s(\sqrt{s^2+1}-s)$ to 1 in \eqref{eq_non_1} and \eqref{eq_non_5}, \ie by considering the bulk value of the thermal conductivity. In this case, the problem \eqref{eq_non_1}-\eqref{eq_non_5} is reduced to an initial value problem via the similarity transformation $\eta = x/\sqrt{t}$. The resulting system has an exact solution, whose temperature and position of the moving boundary are given by 
\begin{align}\label{Neu}
T(x,t) = -1+\frac{\text{erf}\left(\frac{x}{2\sqrt{t}}\right)}{\text{erf}(\lambda)}\,,\quad s(t) = 2\lambda \sqrt{t}\,.  
\end{align}
where $\text{erf}\left(z\right) = \frac{2}{\sqrt{\pi}}\int_0^z e^{-y^2}dy$ is the error function. The constant $\lambda$ is found by solving the transcendental equation 
\begin{align}\label{Neu2}
\beta \sqrt{\pi} \lambda \,\text{erf}(\lambda)\, e^{\lambda^2} = 1\,.  
\end{align}
System \eqref{Neu}-\eqref{Neu2} is referred to as the Neumann solution \cite{Crank}.

\section{Perturbation solution}

The complexity introduced by the size-dependent thermal conductivity in the governing equation prevents an exact similarity solution as in the classical Stefan problem. To make analytical progress a perturbation solution in the limit of large Stefan number is developed. Physically, a large Stefan number corresponds to slow solidification (as can be seen from $\beta\propto 1/(T_f-T_c)$, so large $\beta$ implies small temperature drop). This is consistent with the physical properties of the reference material in Table \ref{table1}, for which $\beta$ is always larger than 1 (even $T_c = 0$ gives $\beta = 1.5$). 

Rescaling time by $t = \beta \hat{t}$ and defining the small parameter $\delta = \beta^{-1}$, the problem becomes 
\begin{subequations}
\begin{align}\label{eq_pert}
\delta\,\frac{\partial T}{\partial \hat{t}} &= 2s\left(\sqrt{s^2+1}-s\right) \frac{\partial^2 T}{\partial z^2} \qquad \mbox{on} \qquad 0<z<s(t) \,,\\ 
T(0,\hat{t}) &= -1\,,\\
T(s(\hat{t}),\hat{t}) &= 0\,,\\ 
\frac{ds}{d\hat{t}} &= 2s\left(\sqrt{s^2+1}-s\right) \left.\frac{\partial T}{\partial z}\right|_{z=s} \,,\label{eq_pert_4}\\
\quad s(0)&=0\,, \label{eq_pert_vis}
\end{align}
\end{subequations}
suggesting an expansion in the form $T(x,\hat{t}) = T_0 + \delta\, T_1 + \mathcal{O} \left(\delta^2\right)$. We obtain the leading and first order problems
\begin{align*}
\mathcal{O}\left(1\right):&& 
0 = \frac{\partial^2 T_0}{\partial z^2}\,, \quad &&T_0(1,\hat{t}) = -1\,, \quad &&T_0(s(\hat{t}),\hat{t}) = 0\,, \\
\mathcal{O}\left(\delta\right):&& 
\frac{\partial T_0}{\partial \hat{t}} = 2s\left(\sqrt{s^2+1}-s\right) \frac{\partial^2 T_1}{\partial z^2}\,, \quad &&T_1(1,\hat{t}) = 0\,, \quad &&T_1(s(\hat{t}),\hat{t}) = 0\,,
\end{align*}
with solutions 
\begin{align*}
T_0 = -1+\frac{x}{s}\,, \qquad T_1 = \frac{s_{\hat{t}}}{12s^3\left(\sqrt{s^2+1}-s\right)}x(s^2-x^2)\,.
\end{align*}
These lead to the temperature profile
\begin{align}
T(x,\hat{t}) &= -1+\frac{x}{s} + \frac{1}{\beta } \frac{s_{\hat{t}}}{12s^3\left(\sqrt{s^2+1}-s\right)}x(s^2-x^2) + \mathcal{O}\left(\delta^2\right)\,. 
\end{align}
Inserting $T \approx T_0 + \delta\, T_1$ into \eqref{eq_pert_4} we obtain the following expression for the speed of the moving boundary
\begin{align}
s_{\hat{t}} = \frac{6\beta}{(1+3\beta)}\left(\sqrt{s^2+1}-s\right)\,,   
\end{align}
which can be readily integrated to give
\begin{align}
s^2 + s\sqrt{s^2+1} + \arcsin(s) =  \frac{12\beta}{(1+3\beta)} \hat{t}\,, 
\end{align}
where the initial condition \eqref{eq_pert_vis} was applied.

\section{Boundary immobilisation and numerical solution}

A typical problem when seeking numerical solutions to moving boundary problems is how to deal with the discretization of a domain whose size changes in time. A way to overcome this difficulty consists in defining a new space variable that transforms the variable domain into a fixed domain. This method is referred to as the boundary immobilization method. 
To immobilise the boundary $s(t)$ we map the space variable $x$ to the unit domain via the Landau-type transformation $\xi = x/s(t)$. Problem \eqref{eq_non_1}-\eqref{eq_non_5} then becomes 
\begin{subequations}
\begin{align}
s\frac{\partial T}{\partial t} &= s_t \xi \frac{\partial T }{\partial \xi} + 2\left(\sqrt{s^2+1}-s\right)\frac{\partial^2 T}{\partial \xi^2}\,  \qquad \mbox{on} \qquad 0<\xi<1 \label{fixedeq}\\
T(0,t) &= -1\,,\\
T(1,t) &= 0\,, \\ 
\beta s_t &= 2\left(\sqrt{s^2+1}-s\right) \left.\frac{\partial T}{\partial \xi }\right|_{\xi=1} \,, \label{bcfixed} \\
s(0)&=0\,. 
\end{align}
\end{subequations}

A Backward Euler semi-implicit finite difference scheme is used on \eqref{fixedeq}, discretising implicitly for $T(\xi,t)$ and explicitly for $s(t)$ and $s_t(t)$, and using second order central differences in space \cite{Cal}. The semi-implicit scheme allows equation \eqref{fixedeq} (containing time dependent coefficients) to be formulated, after discretising, as a matrix linear system which can be solved by inverting the matrix of the system at each time step. The position of the solidification front is found via \eqref{bcfixed}, using a backward difference for $s_t$ and a one-sided second order difference for the partial derivative.

\section{Small time limit} 

A recurring complication associated with the numerical solution of Stefan problems is how to initiate the computation in a region which initially has zero thickness. A typical approach to tackle this issue is to find a small time asymptotic approximation for $T$ and $s$ and use this as the initial condition in the numerical scheme \cite{mit09}. 

To study the small time limit we assume $s(t)$ to be a power of time that matches the initial condition $s(0)=0$. Hence, we set $s=\lambda_2 t^{p}$ where $\lambda_2$ and $p$ are constants. We rescale time as $t = \varepsilon \tau$, where $\epsilon\ll 1$, to obtain the expressions $s=\lambda_2 \varepsilon^{p} \tau^{p}$ and $s_t=p\lambda_2\varepsilon^{p-1}\tau^{p-1}$ for the position and velocity of the solidification front, respectively. Substituting $s$ and $s_t$ into \eqref{bcfixed} leads to
\begin{align*}
\frac{\beta}{2} p\lambda_2\varepsilon^{p-1}\tau^{p-1} = \left(\sqrt{\varepsilon^{2p}\tau^{2p}+1}-\lambda_2\varepsilon^{p}\tau^{p}\right) \left.\frac{\partial T}{\partial \xi }\right|_{\xi=1}\,, 
\end{align*}
which, after multiplying the right hand side by $(\sqrt{\varepsilon^{2p}\tau^{2p}+1}+\lambda_2\varepsilon^{p}\tau^{p})(\sqrt{\varepsilon^{2p}\tau^{2p}+1}+\lambda_2\varepsilon^{p}\tau^{p})^{-1}$, rearranging and noting that $\sqrt{\varepsilon^{2p}\tau^{2p}+1}\approx 1$ for $\varepsilon^{2p}\tau^{2p}\ll1$, gives 
\begin{align}\label{balance}
\frac{\beta}{2} p\lambda_2\varepsilon^{p-1}\tau^{p-1} \left( 1 +\lambda_2\varepsilon^{p}\tau^{p}\right) \approx \left.\frac{\partial T}{\partial \xi }\right|_{\xi=1}\,. 
\end{align}

Now we need to choose $p$. We realize that, in order for the front to move, the left hand side of \eqref{balance} must be $\mathcal{O}(1)$, thereby balancing the right hand side. The balance is only satisfied for $p=1$, so we obtain
\begin{align}\label{eqfors}
s = \lambda_2 t\,,
\end{align}
for $t\rightarrow0$. Substituting \eqref{eqfors} into \eqref{fixedeq} and taking the limit  $t\rightarrow0$ gives
\begin{align*}\label{}
0 =  \xi\frac{\lambda_2}{2}\frac{\partial T }{\partial \xi} + \frac{\partial^2 T}{\partial \xi^2}\,,
\end{align*}
with solution 
\begin{align}\label{solsmallt}
T = -1 + \frac{\text{erf}\,\left(\frac{\sqrt{\lambda_2}}{2}\xi \right)}{\text{erf}\,\left(\frac{\sqrt{\lambda_2}}{2}\right)}\,. 
\end{align}
Finally, the constant $\lambda_2$ is found by substituting \eqref{eqfors} and \eqref{solsmallt} into \eqref{bcfixed} and taking the limit $t\rightarrow0$, leading to the transcendental equation 
\begin{align}\label{tran2}
\frac{\beta}{2}\sqrt{\pi} \sqrt{\lambda_2}\, \text{erf}\,\left(\frac{1}{2}\sqrt{\lambda_2}\right)e^{\lambda_2/4}=1\,. 
\end{align}
From a physical standpoint, expression \eqref{eqfors} states that the speed of the solidification front, $s_t$, is constant as $t\rightarrow0$, in contrast to the Neumann solution \eqref{Neu2} which gives $s_t\rightarrow\infty$ as $t\rightarrow0$.    

Next, backward-substituting $\xi = x/s(t)$ into \eqref{solsmallt} and using \eqref{eqfors} we obtain an initial size and temperature profile for the solid phase at some small time $t>0$, which we use to initialise our numerical scheme. We note that by defining $\sqrt{\lambda_2}/2 = \lambda$ the transcendental equation \eqref{tran2} becomes equal to that in \eqref{Neu2}.

\section{Results and discussion}

\begin{figure}[h!]
\centering
\includegraphics[width=0.45\textwidth]{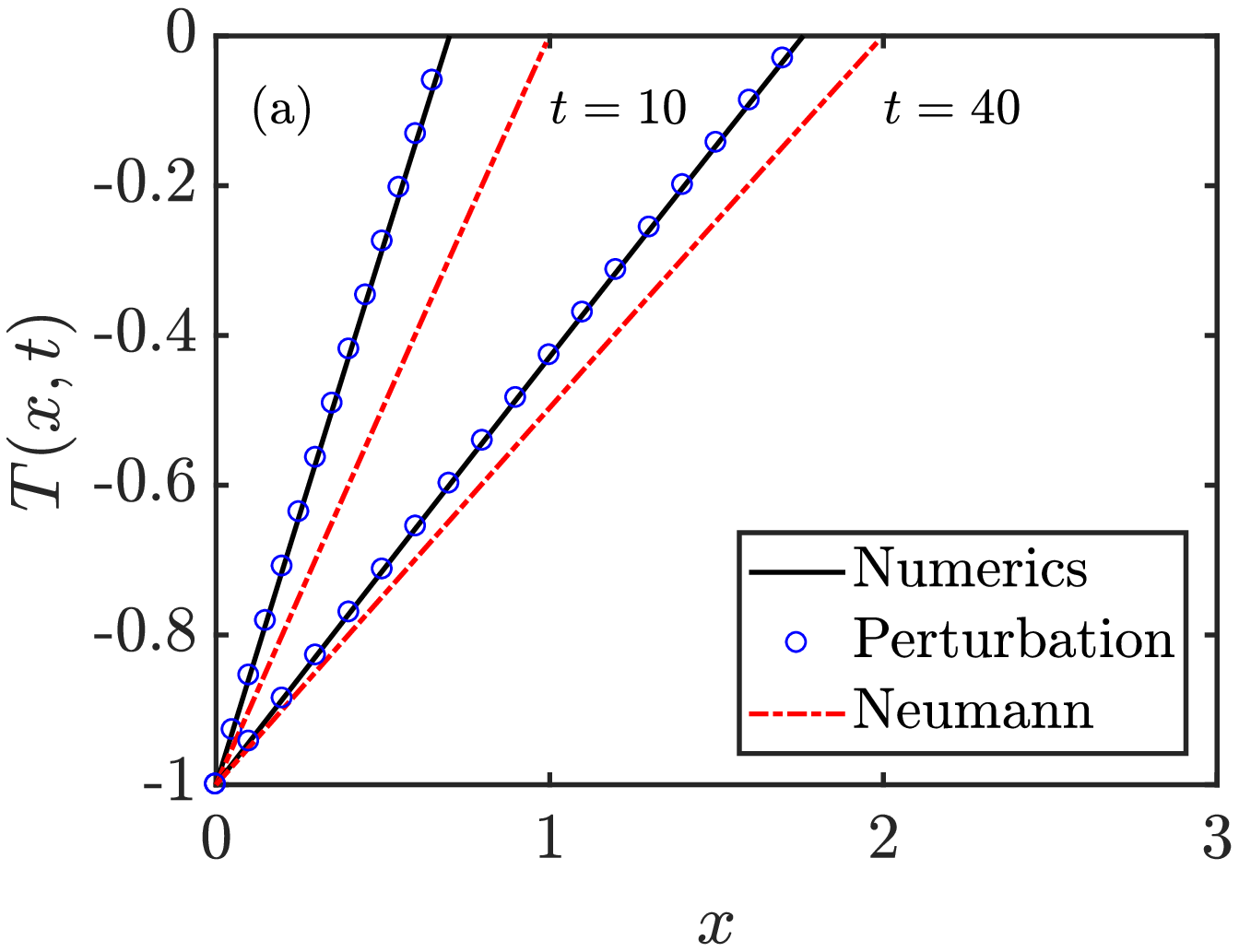}\includegraphics[width=0.45\textwidth]{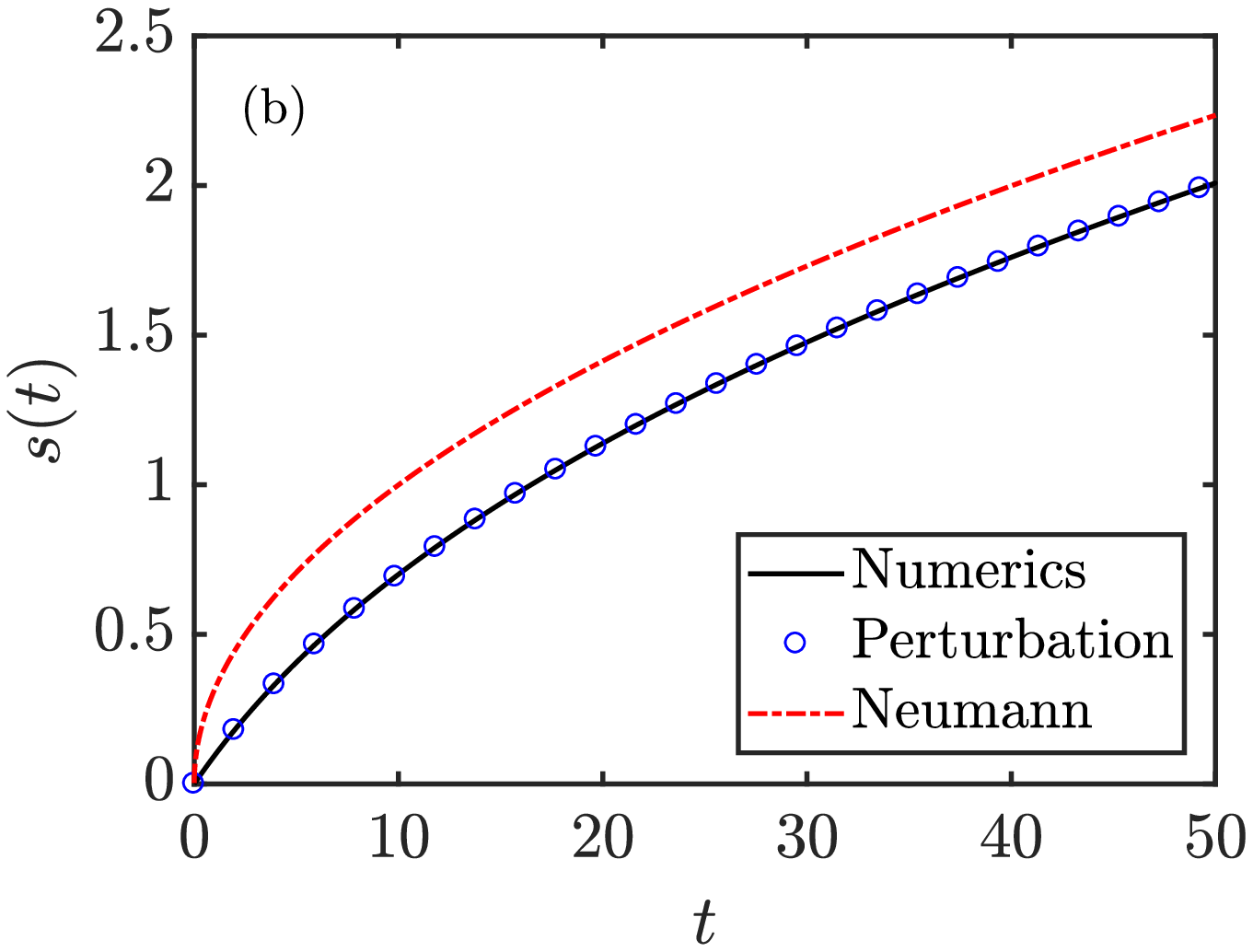}
\includegraphics[width=0.45\textwidth]{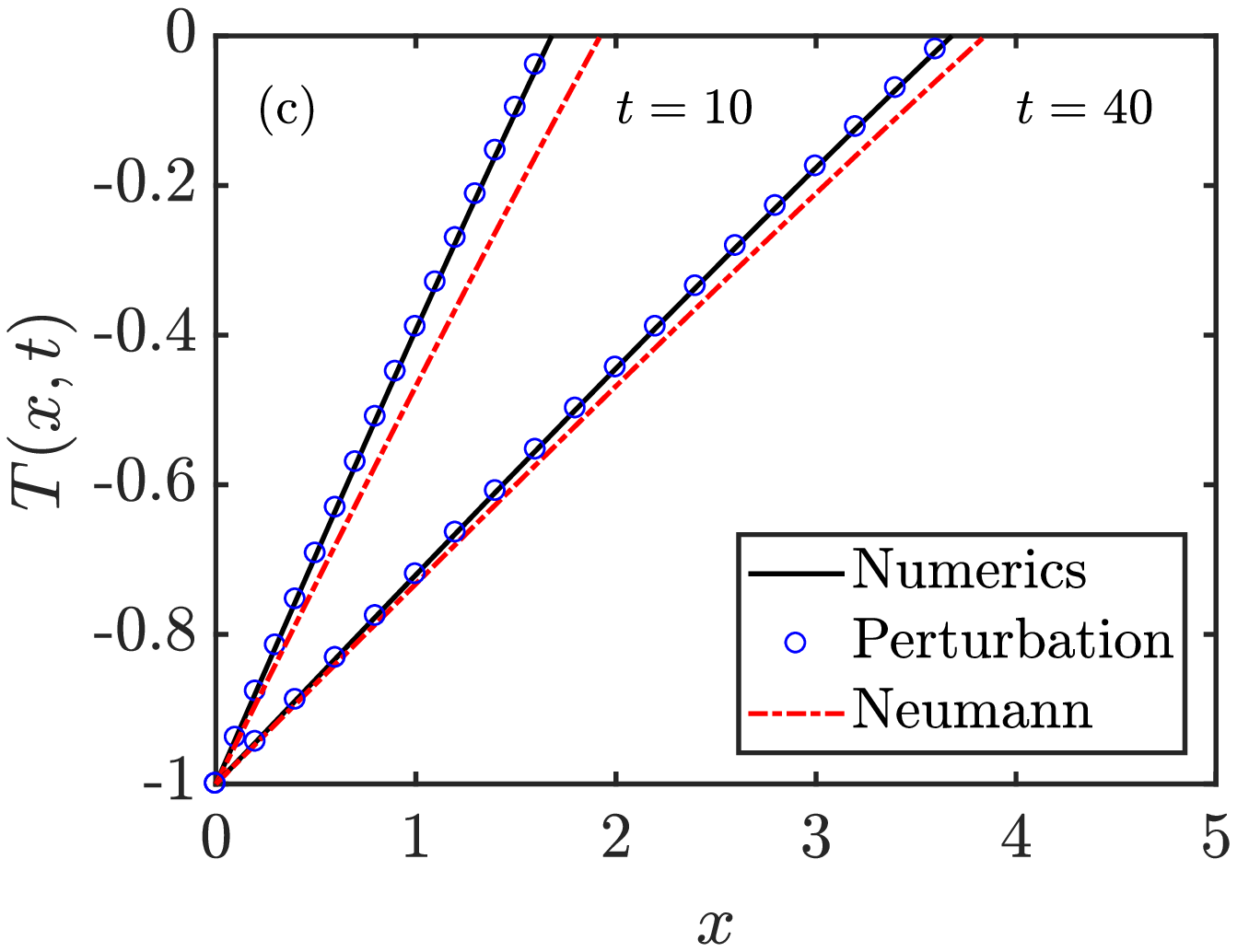}\includegraphics[width=0.45\textwidth]{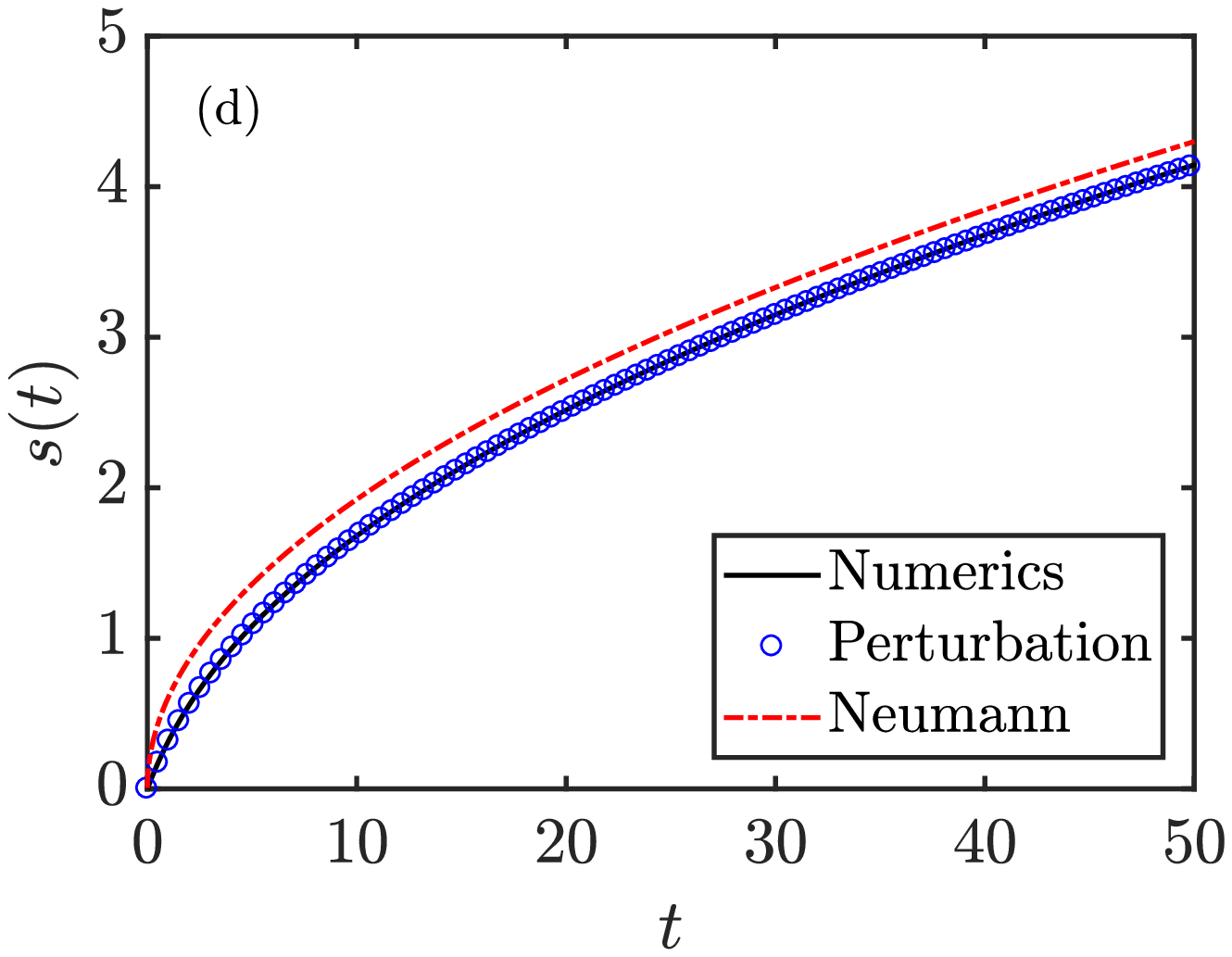}
\caption{Temperature profiles at different times and position of the solidification front for the Stefan problem with size-dependent thermal conductivity and the standard Stefan problem. The circles, the solid line and the  dash-dotted line represent the perturbation solution, the numerical solution and the Neumann solution, respectively. Panels (a)-(b) correspond to $\beta = 20$ and (c)-(d) to $\beta = 5$. }
\label{results1}
\end{figure}

In Figure~\ref{results1} we present the perturbation and numerical solution to the Stefan problem with size-dependent thermal conductivity for two values of the Stefan number, $\beta = 20$ and $\beta = 5$, along with the corresponding Neumann solution to the classical problem. The perturbation and numerical solutions to the problem with variable thermal conductivity show excellent agreement, even for a relatively small Stefan number $\beta = 5$, thereby validating the accuracy of the numerical solution. 

\begin{figure}[h!]
\centering
\includegraphics[width=0.47\textwidth]{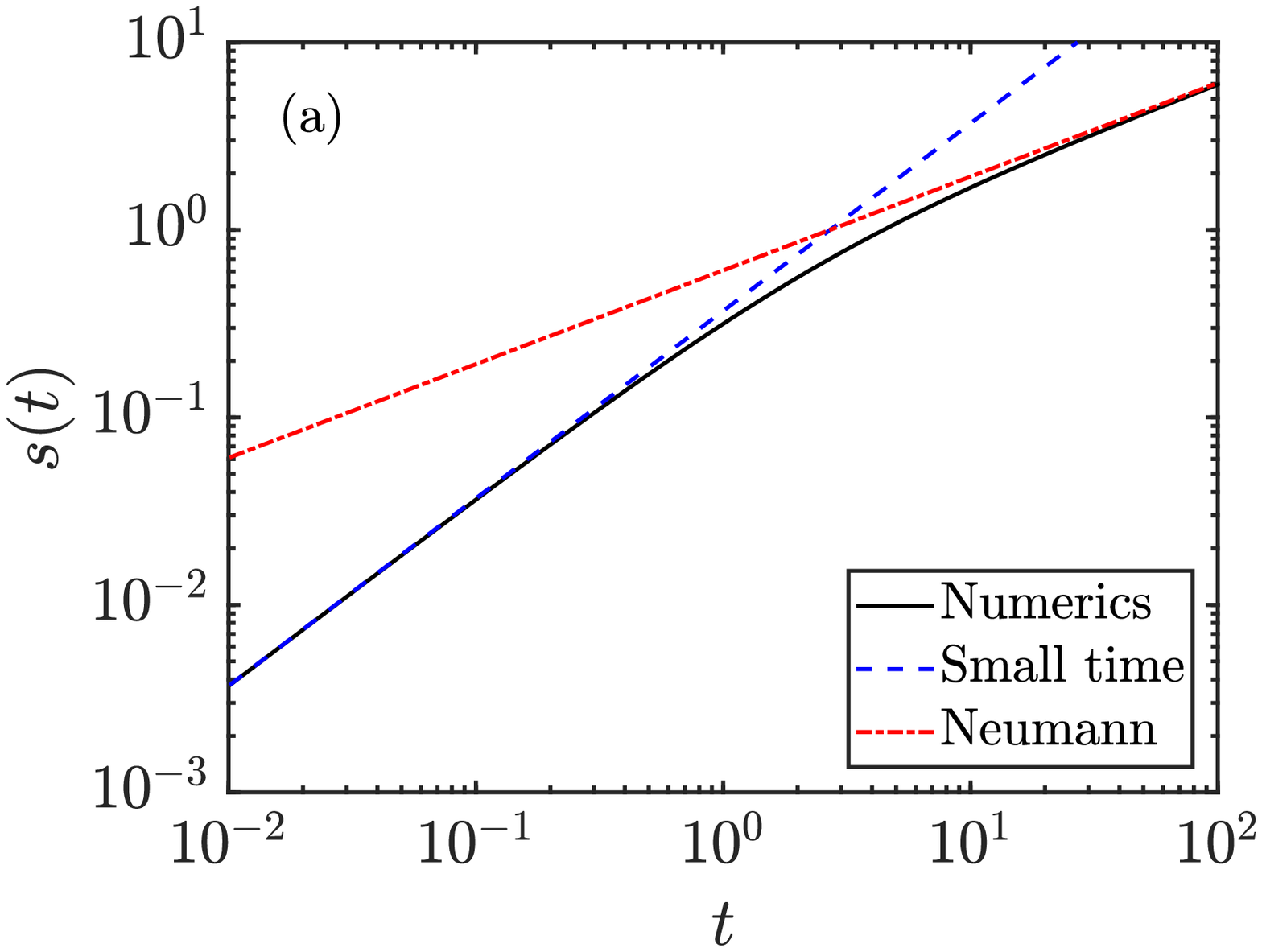}\includegraphics[width=0.47\textwidth]{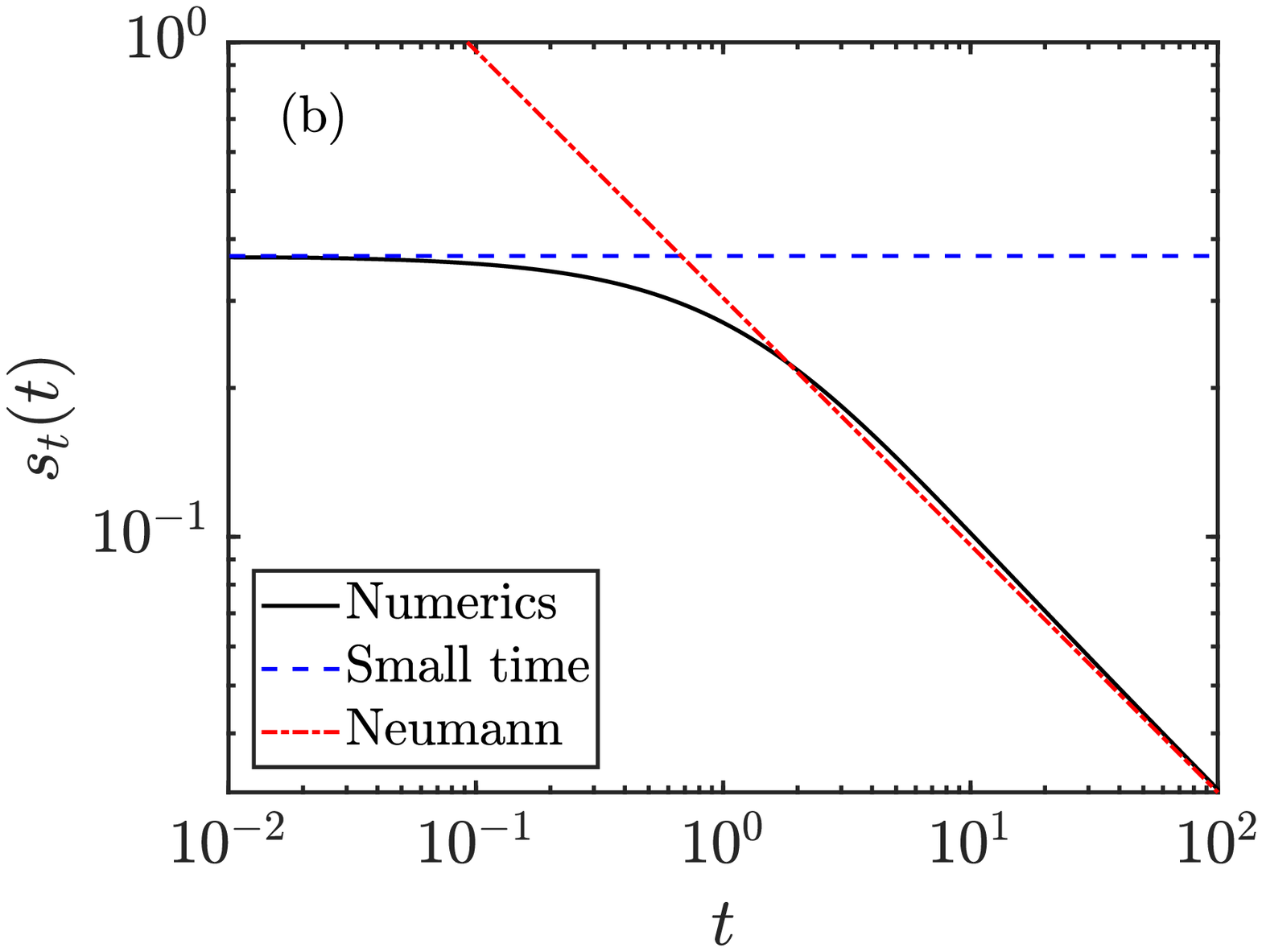}
\caption{Position and velocity of the solidification front as a function of time spanning several orders of magnitude for the case $\beta= 5$. The solid line, dashed line and the dash-dotted line represent the numerical, the small time asymptotic and the Neumann solutions, respectively. }
\label{results2}
\end{figure}

By comparing the solutions of the variable conductivity problem with the Neumann solution we can assert that the overall effect of the size-dependent thermal conductivity on the solidification process is to delay the propagation of the temperature and the solidification front. Indeed, the delay is caused by a small value of the thermal conductivity, represented by the term $2s(\sqrt{s^2+1}-s)$, in the early stages of the solidification process when $s\sim\mathcal{O}(1)$. However, the value of $2s(\sqrt{s^2+1}-s)$ quickly converges to the maximum value of 1 (for instance, $s\approx2$ already gives $2s(\sqrt{s^2+1}-s)\approx0.95$) and the solution to the Stefan problem with size-dependent thermal conductivity converges to the Neumann solution. This is further illustrated in Figure~\ref{results2}.

In Figure \ref{results2} we show the position ((a) panel) and velocity ((b) panel) of the solidification front spanning several orders of magnitude in time for the case $\beta = 5$. The solution of the Stefan problem with size-dependent thermal conductivity represented by the solid line (for clarity we only show the numerical solution), clearly shows different qualitative behaviour for small and large times. For small times, the position of the front behaves linearly with time (according to eq. \eqref{eqfors}) which is represented by the dashed line which has a slope 1. For large times, the position is well approximated by the Neumann solution, which is proportional to $\sqrt{t}$ (according to eq. \eqref{Neu2}), and so gives a slope of 0.5. The fact that $s \propto t$ as $t\rightarrow0$ has an important physical implication, which is that the speed of the solidification front is constant and, thus, finite when the solidification begins, as indicated by te small time asymptotic limit in panel (b). In contrast, the Neumann solution gives $s_t\propto1/\sqrt{t}$, leading to an infinite speed for the solidification front at the beginning of the process. 

\section{Conclusions}

In this paper we analysed a one-phase Stefan problem subject to  size-dependent thermal conductivity. The results indicate that the size-dependent thermal conductivity induces a delay in the propagation of the solidification front, which is caused by the reduced value of the conductivity when the solid phase is small.   
A small time limit analysis revealed that the speed of the solidification front is constant as $t\rightarrow0$, whereas the Neumann solution to the classical problem predicts an unrealistic infinite velocity as time goes to zero. This result is very relevant as it gives a physically realistic mathematical description of how solidification phenomena initiates. When the size of the solid phase is approximately order 1, the qualitative behaviour of the solution switches to the standard $\sqrt{t}$ proportionality and the solution to the problem tends to the classical Neumann solution.

\section*{Acknowledgements}

This work was partially funded by the Mathematics for Industry Network (MI-NET, COST Action TD1409). The author would like to thank Vincent Cregan for a critical reading of the manuscript.


\bibliographystyle{elsarticle-num} 
\bibliography{biblio}

\end{document}